\documentclass{amsart}

\usepackage[T1]{fontenc}
\usepackage[utf8]{inputenc}
\usepackage{booktabs}
\usepackage{siunitx}
\usepackage{pgfplots}
\usepackage{floatrow}
\usepackage{caption}
\usepackage{microtype}
\usepackage{varioref}
\usepackage[autostyle]{csquotes}
\usepackage[backend=biber,safeinputenc,sorting=nyt,maxbibnames=10 ]{biblatex}
\usepackage[colorlinks=true,linkcolor=blue,urlcolor=violet,citecolor=red]{hyperref}

\usepackage[utf8]{inputenc}	
\usepackage[T1]{fontenc}	
\usepackage{lmodern}

\usepackage{amssymb}
\usepackage{amsbsy}
\usepackage{amscd}
\usepackage{amsmath}
\usepackage{amsthm}
\usepackage{amsxtra}
\usepackage{mathtools}
\usepackage[new]{old-arrows}

\usepackage{mathtools}		
\usepackage{amssymb}
\usepackage{systeme}		

\usepackage{siunitx}		

\usepackage{textcomp}		

\usepackage[shortlabels]{enumitem} 

\usepackage{xspace}			

\usepackage{xcolor}			

\usepackage{xifthen} 		

\usepackage{dsfont}
\usepackage{stmaryrd}
\usepackage{mathrsfs}

\usepackage{graphicx}		
\usepackage{epsfig}			

\usepackage{tikz}			
\usetikzlibrary{cd}
\usepackage{tkz-tab}
\usetikzlibrary{matrix}
\usepackage{dynkin-diagrams}

\usepackage{caption}
\usepackage{subcaption}		

\usepackage{listings}		

\usepackage{varioref}
\usepackage[autostyle]{csquotes}
\usepackage[backend=biber,safeinputenc,sorting=nyt]{biblatex}
\usepackage[colorlinks=true,linkcolor=blue,urlcolor=violet,citecolor=red]{hyperref}

\allowdisplaybreaks[1] 		

\usepackage{marginnote}
\newcommand{\mc}{\mathcal}

\renewcommand{\leq}{\leqslant}
\renewcommand{\geq}{\geqslant}

\renewcommand{\leq}{\leqslant}
\renewcommand{\geq}{\geqslant}
\newcommand{\+}{\oplus}
\renewcommand{\*}{\otimes}

\newcommand{\Z}{\mathbb{Z}}
\newcommand{\C}{\mathbb{C}}
\newcommand{\R}{\mathbb{R}}

\newcommand{\g}{\mathfrak{g}}

\renewcommand{\sl}{\mathfrak{sl}}

\renewcommand{\sp}{\mathfrak{sp}}

\newcommand{\ens}[2]{\ifthenelse{\equal{#2}{pp}}{#1_{>0}}{\ifthenelse{\equal{#2}{p}}{#1_{\geq0}}{\ifthenelse{\equal{#2}{nn}}{#1_{<0}}{\ifthenelse{\equal{#2}{n}}{#1_{\leq0}}{#1_{#2}}}}}}
\newcommand{\VA}{\ensuremath{V^{k}(\g)}}

\newcommand{\W}{\ensuremath{\mathcal{W}}}
\newcommand{\WA}{\ensuremath{\W^k(\g,f)}}
\newcommand{\SWA}{\ensuremath{\W_k(\g,f)}}

\renewcommand{\O}{\mathbb{O}}

\newcommand{\Imax}{\mathcal{I}_k}

\newcommand{\fmin}{\ensuremath{f_\mathrm{min}}}
\newcommand{\subreg}{\mathrm{subreg}}
\newcommand{\reg}{\mathrm{reg}}

\newcommand{\NO}[1]{\ensuremath{:#1:}}
\newcommand{\OPE}[3]{\ensuremath{\frac{#2}{#3(z-w)^{#1}}}}
\newcommand{\charge}{\ensuremath{c_k}}

\newcommand{\vacuum}{|0\rangle}

\DeclareMathOperator{\Vir}{Vir}

\theoremstyle{theorem}
\newtheorem{Theorem}{Theorem}[section]

\theoremstyle{nonumberplain}

\theoremstyle{plain}
\newtheorem{Proposition}[Theorem]{Proposition}

\newtheorem{Corollary}[Theorem]{Corollary}

\newtheorem{Conjecture}[Theorem]{Conjecture}

\theoremstyle{remark}

\newtheorem{Remark}[Theorem]{Remark}

\newenvironment{Proof}{\proof}{\endproof}

\lstnewenvironment{codeGAP}{\lstset{language=GAP,basicstyle=\ttfamily}}{}
\lstnewenvironment{codeGAPboxed}{\lstset{language=GAP,basicstyle=\ttfamily,frame=single,framexleftmargin=5mm,numbers=left,numberstyle=\scriptsize,numbersep=5pt}}{}

\newcolumntype{M}[1]{>{\centering\arraybackslash}m{#1}}

\title[OPEs for $\W$-algebras of rank $2$]{\small OPEs of rank two $\W$-algebras}
\author{Justine Fasquel}
\address{School of Mathematics and Statistics, University of Melbourne, Parkville, VIC 3010, Australia.}
\email{justine.fasquel@unimelb.edu.au}

\begin{document}
\begin{abstract}
	In this short note, we provide OPEs for several affine $\W$-algebras associated with Lie algebras of rank two and give some direct applications. 
\end{abstract}

\maketitle

\section{Introduction}

The $\W$-algebra $\WA$ is obtained from an Hamiltonian reduction of the affine vertex algebra $\VA$ \cite{FF90,KRW03}. Hence, both structure are closely related.
However, if the description of $\VA$ in term of strong generators and relations (OPEs) is well-known (see for instance \cite{FB04,Kac17}), it is no longer the case for the $\W$-algebras. 
Indeed, the latter do not directly come from Lie algebras and we need the formalism of vertex algebras to describe them.
The computation of OPEs is difficult in general and only few examples were computed explicitely \cite{Ara13,CL18,Zam85}.

In this short note, we provide the OPEs for $\W$-algebras associated with Lie algebras of rank two (i.e. $\sl_3$, $\sp_4$ and $G_2$). 

The OPEs consist in a powerful tool to prove important and precise results. For instance, they can be used to prove rationality of $\W$-algebras \cite{Ara13,Fas22}, compute associated varieties \cite{Fasthesis}, or find isomorphisms between vertex algebras \cite{AKM+17,AKM+18,MPS22}.



The present paper is organized as follows. In Sect.~\ref{sec:construction} we recall the construction of the $\W$-algebra due to Kac, Roan, Wakimoto \cite{KRW03}. 
Sect.~\ref{sec:OPEs} presents the OPEs of $\W$-algebras associated to a Lie algebra $\g$ of rank two.
We use these OPEs to emphasize isomorphisms of $\W$-algebras for particular values of the level $k$. These levels are listed in Sect.~\ref{sec:collapsing_levels}.

\subsection*{Acknowledgments}
This paper is based on the author's Ph.D. thesis \cite{Fasthesis}.
The author would like to thank her advisors A. Moreau and T. Arakawa for suggesting the problem.
She is very grateful to D. Adamovi\'c for useful discussions on collapsing levels.
She also thanks A. Milas for the interesting feedbacks and exchanges.

\noindent The author acknowledges support from the Labex CEMPI (ANR-11-LABX-0007-01). Her research is supported by a University of Melbourne Establishment Grant.


\section{Structure of affine $\W$-algebras}\label{sec:construction}

We assume the lector familiar with the formalism of vertex algebras 
and we refer to \cite{KRW03} and \cite{Ara05,Fasthesis,KW04} for details on the construction of $\W$-algebras.


Let $\g$ be a complex simple finite dimensional Lie algebra.
Consider $f$ a nilpotent element of $\g$ and $k$ a complex number. The (affine) $\W$-algebra $\WA$ is the vertex algebra obtained as the zero-th cohomology of a complex $\mathscr{C}(\g,f,k)$ constructed from the affine vertex algebra $\VA$. It is a particular case of BRST reduction usually referred as the Drinfeld-Sokolov reduction of $\VA$. 
We write
$$\WA=H_f^0(\VA).$$
Its simple graded quotient is denoted by $\SWA$. 
Provided that $k$ is non-critical, i.e. $k\neq -h^\vee$, $\WA$ inherits a structure of conformal vertex algebra from those of $\VA$.
Denote $L(z)$ the usual conformal field defined in \cite{KRW03}.

Let embed $f$ into an $\sl_{2}$-triple $(e,h,f)$ given by the Jacobson-Morosov theorem. We have
\begin{equation*}
	[h,e]=2e,\qquad [h,f]=-2f,\qquad\text{and}\qquad [e,f]=h.
\end{equation*}
The semisimple element $x_0:=h/2$ induces an $\frac{1}{2}\Z$-gradation on $\g$,
\begin{equation}\label{eq:decompg}
\g = \bigoplus_{j \in \frac{1}{2}\Z} \g_j,
\end{equation}
where $\g_j=\{y \in\g\mid [x_0,y ] = j y \}$. 
To any $a \in \g_{-j}$, $j\in\frac{1}{2}\Z$, we can associate a field $J^a(z)$ of $\mathscr{C}(\g,f,k)$. This field has conformal weight $1+j$ with respect to $L$.
The fields $\{J^a\}_{a\in\g}$ play an important role in the structure of the $\W$-algebra $\WA$.
Let $\g^f$ be the centralizer of $f$ in $\g$, and set $\g^f_j:=\g^f\cap\g_j$ for all $j \in \frac{1}{2}\Z$. By the theory of $\sl_2$, we have 
\begin{align}\label{eq:gradgf} 
	\g^f=\bigoplus_{j\leq0}\g^f_j.
\end{align}
The strong generators of the $\W$-algebra $\WA$ are in one to one correspondance with the elements of a basis of $\g^f$ with respect to the grading \eqref{eq:gradgf}.

\begin{Theorem}[\cite{KW04}]
	\label{thm:Wstructure}
	For each $a\in\g^f_{-j}$, $j\geq 0$, there exists a field $J^{\{a\}}(z)$ of $\WA$ of conformal weight $1+j$ 
	with respect to $L$
	such that $J^{\{a\}}(z)-J^{a}(z)$ is a linear combination of normally order products of the fields 
	$J^b(z)$, where $b \in \g_{-s}$, $0 \leq s <j$ and their derivatives. 
	
	Let $\{a_i\}_{i\in I}$ be a basis of $\g^f$ compatible with the graduation \eqref{eq:gradgf}. 
	Then $\WA$ is freely strongly generated by the fields $\{J^{\{a_i\}}\}_{i\in I}$.
\end{Theorem}

In practice, to construct $J^{\{a\}}(z)$ from $J^{a}(z)$, 
we write a linear combination of fields as in the theorem and find coefficients so that the field $J^{\{a\}}(z)$ is $d_{(0)}$-closed.
When $f$ is an even nilpotent element there exists an algorithm to construct the strong generators of the $\W$-algebra $\WA$ (see \cite{Fasthesis,KW04}).
In \cite{KW04}, Kac and Wakimoto give formulas to construct strong generators with conformal weights $1$, $1/2$ and $0$, and they compute OPEs between them. 
This provides in particular an explicit description of all $\W$-algebras associated with minimal nilpotent elements. 
Indeed, if $f$ belongs to the minimal nilpotent orbit $\O_{\min}$ then the grading \eqref{eq:decompg} induced by $x$ is \emph{minimal}:
$$\g=\g_{-1}\oplus\g_{-\frac{1}{2}}\oplus\g_0\oplus\g_{\frac{1}{2}}\oplus\g_1,$$
with $\g_{-1}=\C f$ and $\g_1=\C e$.

\section{OPEs of $\W$-algebras associated with rank two Lie algebras}\label{sec:OPEs}

In this section, we present OPEs of the $\W$-algebra $\WA$ where $\g$ is a simple Lie algebra of rank two and $k\neq-h^\vee$ is a non critical level. 
For simplicity, $L$ will always denote the conformal vector.
In the cases we consider, we were able to find strong generators which are all -- except $L$ -- \emph{primary} with respect to the conformal structure induced by $L$. It means that for any strong generator $J\neq L$ of conformal weight $\Delta$,
$$L(z)J(w)\sim\OPE{2}{\Delta}{}J(w)+\OPE{}{1}{}\partial J(w).$$
We take these primary generators in the following. With this choice, the OPEs involving $L$ are uniquely determined. We omit them in the computations.

\subsection{$\W$-algebras associated with nilpotent elements of $\sl_3$}\label{subsec:OPEsA2}

The simple Lie algebra $\sl_3$ of type $A_2$ admits two non-trivial nilpotent orbits: the principal and the minimal one.

The corresponding $\W$-algebras are among the first examples of $\W$-algebras which have been described explicitly. They are respectively called the Bershadsky-Polyakov vertex algebra \cite{Ber91,Pol90} and the Zamolodchikov vertex algebra \cite{Zam85}.



In \cite{Ara13}, Arakawa gives an explicit description of the Bershadsky-Polyakov vertex algebra which corresponds to the $\W$-algebra associated with a minimum nilpotent element of $\sl_3$. 
The $\W$-algebra $\W^k(\sl_3,\fmin)$ is strongly generated by the fields $J(z)$, $G^\pm(z)$ and $L(z)$ of conformal weights $1$, $\frac{3}{2}$ and $2$ respectively.
These fields satisfy the following relations:
\begin{align*}
J(z)J(w)&\sim\OPE{2}{(3+2k)}{},\\
J(z)G^\pm(w)&\sim\pm\OPE{}{1}{}G^\pm(w),\\
%
%
%
%
G^\pm(z)G^\pm(w)&\sim0,\\
G^+(z)G^-(w)&\sim-\OPE{3}{(1+k)(3+2k)}{}+\OPE{2}{3(1+k)}{}J(w)\\
+\OPE{}{1}{}&\left(-(3+k)L(w)+3\NO{J(w)^2}-\frac{3(1+k)}{2}\partial J(w)\right),
\end{align*}
where 
\[\charge=-\frac{(1+3k)(3+2k)}{3+k}.\]



On another side, the $\W$-algebra $\W^k(\sl_3)$ is strongly generated by the fields $L(z)$ and $W(z)$ of conformal weights $2$ and $4$ respectively satisfying the OPE:
\begin{align*}
%
%
W(z)W(w)&\sim\OPE{6}{w_kc_k}{3}+\OPE{4}{2w_k}{}L(w)+\OPE{3}{w_k}{2}\partial L(w)\\
+\OPE{2}{1}{}&\left(\frac{2(3+k)^2}{3}\NO{L(w)^2}-\frac{3(3+k)^2(2+k)^2}{4}\partial^2L(w)\right)\\
+\OPE{}{1}{}&\left(\frac{2(3+k)^3}{3}\NO{L(w)\partial L(w)}-\frac{(3+k)^2(18+14k+3k^2)}{18}\partial^3L(w)\right),
\end{align*}
where 
\[\charge=-\frac{2(5+3k)(9+4k)}{3+k},\]
and
$$w_k=-\frac{(3+k)^2(4+3k)(12+5k)}{6}.$$

\subsection{$\W$-algebras associated with nilpotent elements of $\sp_4$}\label{subsec:OPEsB2}

There are three non-trivial nilpotent orbits in $\sp_4$ corresponding to the principal, the subregular, the minimal nilpotent orbits.




The minimal $\W$-algebra $\W^k(\sp_4,\fmin)$ is of type $(1^3,\frac{3}{2}^2,2)$, strongly generated by the fields $J(z)$, $F^\pm(z)$, $G^\pm(z)$ and $L(z)$. They satisfy the following OPEs:
\begin{align*}
	J(z)J(w)&\sim\OPE{2}{(1+2k)}{},\\
	J(z)F^\pm(w)&\sim\pm\OPE{}{2}{}F^\pm(w),\\
	F^\pm(z)F^\pm(w)&\sim0,\\
	F^+(z)F^-(w)&\sim\OPE{2}{(1+2k)}{2}+\OPE{}{1}{}J(w),\\
	J(z)G^\pm(w)&\sim\pm\OPE{}{1}{}G^\pm(w),\\
	G^\pm(z)F^\pm(w)&\sim0,\\
	G^\pm(z)F^\mp(w)&\sim\OPE{}{1}{}G^\mp(w),\\
	%
	%
	G^\pm(z)G^\pm(w)&\sim\pm\OPE{2}{4(2+k)}{}F^\pm(w)\pm\OPE{}{2(2+k)}{}\partial F^\pm(w),\\
	G^+(z)G^-(w)&\sim\OPE{3}{2(1+2k)(2+k)}{}+\OPE{2}{2(2+k)}{}J(w)\\+\OPE{}{1}{}&\left(-2(3+k)L(w)+4\NO{F^+(w)F^-(w)}+\NO{J(w)^2}+k\partial J(w)\right),
\end{align*}
where \[\charge=-\frac{3(k+1)(2k+1)}{3+k}.\]


The explicit description of the subregular $\W$-algebra $\W^k(\sp_4,f_{\subreg})$ firstly appeared in \cite{Fas22}. 
The $\W$-algebra $\W^k(\sp_4,f_{\subreg})$ is strongly generated by the fields $J(z)$, $G^\pm(z)$ and $L(z)$ of conformal weights $1$, $2$ and $2$ respectively. They satisfy the OPEs: 
\begin{align*}
J(z)J(w)&\sim\OPE{2}{(2+k)}{},\\
J(z)G^\pm(w)&\sim\pm\OPE{}{1}{}G^\pm(w),\\
%
%
G^\pm(z)G^\pm(w)&\sim0,\\
G^+(z)G^-(w)&\sim-\OPE{4}{3(1+k)(2+k)^2}{}-\OPE{3}{3(1+k)(2+k)}{}J(w)\\
+\OPE{2}{1}{}&\left((2+k)(3+k)L(w)-(3+2k)\NO{J(w)^2}-\frac{3(1+k)(2+k)}{2}\partial J(w)\right)\\
+\OPE{}{1}{}&\left((3+k)\NO{L(w)J(w)}+\frac{(3+k)(2+k)}{2}\partial L(w)-\NO{J(w)^3}\right.\\
&\qquad\left.-(3+2k)\NO{J(w)\partial J(w)}-\frac{(5+4k+k^2)}{2}\partial^2 J(w)\right),
\end{align*}
where 
\[\charge=-\frac{2(9+16k+6k^2)}{3+k}.\]



The prinicipal $\W$-algebra $\W^k(\sp_4)$ is of type $(2,4)$ strongly generated by the fields $L(z)$ and $W(z)$. The field $W(z)$ satisfy the OPE:
\begin{align*}
%
W(z)W(w)&\sim\OPE{8}{w_0c_k}{4}+\OPE{6}{2w_0}{}L(w)+\OPE{5}{w_0}{}\partial L(w)\\
+\OPE{4}{1}{}&\left(2w_2W(w) - w_1\left(21(3+k)\NO{L(w)^2} -
\frac{3(162 + 139 k + 30 k^2)}{2} \partial^2L(w)\right)\right)\\
+\OPE{3}{1}{}&\left(w_2\partial W(w) - w_1 \left(21 (3 + k)\NO{\partial L(w)L(w)} -\frac{(162 + 139 k + 30 k^2)}{3} \partial^3 L(w)\right)\right)\\
+\OPE{2}{(3 + k)}{}&\left(17 + 12 k + 
2 k^2\right) \left(\frac{(684 + 653 k + 150 k^2)}{2} \partial^2 W(w)- 
84 (3 + k)  \NO{L(w)W(w)}\right) \\
+\OPE{2}{w_3}{}&\left(4 (3 + k)^2 (65 + 27 k) (69 + 32 k) \NO{L(w)^3} - (3 + k) w_4 \NO{\partial^2L(w)L(w)} \right.\\&\quad\left.- (3 + k) (540 + 509 k + 118 k^2) (812 + 701 k + 
150 k^2) \NO{\partial L(w)^2} + w_5 \partial^4 L(w)\right)\\
+\OPE{}{(3 + k)}{2}& (17 + 12 k + 2 k^2) \left((162 + 139 k + 30 k^2)\partial^3 W(w)\right.\\
&\quad-84 (3 + k) (\NO{L(w)\partial W(w)} + \NO{\partial L(w)W(w)})\big) \\
+\OPE{}{w_3 }{}&\left(6 (3 + k)^2 (65 + 27 k) (69 + 32 k)\NO{\partial L(w) L(w)^2} \right.\\
&\quad\left.- (3 + k)  \left(w_6 \NO{\partial^2L(w)\partial L(w)}+w_7\NO{\partial^3L(w)L(w)}\right)+ w_8 \partial^5L(w)\right).
\end{align*}
where 
\[\charge=-\frac{2(12+5k)(13+6k)}{3+k},\]
and
\begin{align*}
	w_0 &=(3+k)^2(3+2k)(8+3k)(11+5k)(11+6k)(18+7k)(19+8k)(747+674k+150k^2),\\
	w_1 &= (3 + k)^2 (3 + 2 k) (8 + 3 k) (11 + 5 k) (11 + 6 k) (18 + 
	7 k) (19 + 8 k),\\
	w_2&= 45(3 + k) (3 + 2 k) (8 + 3 k) (17 + 12 k + 2 k^2),\\
	w_3&= \frac{1}{2} (3 + k)^2 (3 + 2 k) (8 + 3 k),\\
	w_4&=566676 + 998131 k + 658239 k^2 + 192662 k^3 + 
	21120 k^4,\\
	w_5&=823635 + 2187341 k + 2413982 k^2 + 1417736 k^3 + 
	467512 k^4 + 82100 k^5 + 6000 k^6\\
	w_6&=3(95769 + 168158 k + 110646 k^2 + 32332 k^3 +3540 k^4),\\
	w_7&=125046 + 220605 k + 145649 k^2 + 42666 k^3 + 4680 k^4,\\
	w_8&=\frac{3}{20}(823635 + 2187341 k + 2413982 k^2 + 1417736 k^3 + 467512 k^4 + 82100 k^5 + 6000 k^6).
\end{align*}

\subsection{$\W$-algebras associated with nilpotent elements of $G_2$}\label{subsec:OPEsG2}

Consider the Lie algebra $G_2$ of rank two.
According to the Bala-Carter classification, $G_2$ has four non-trivial nilpotent orbits: the three canonical nilpotent orbits -- $\O_{\min}$, $\O_{\subreg}$, and $\O_{\reg}$ -- and an additional nilpotent orbit of dimension $8$, denoted $\tilde{A_1}$~\cite[Chap.~8]{CM93}.

The $\W$-algebra $\W^k(G_2,\fmin)$ is of type $(1^3,\frac{3}{2}^4,2)$, strongly generated by the fields $J(z)$, $F^\pm(z)$, $G^\pm(z)$, $W^\pm(z)$, and $L(z)$. They satisfy the OPEs:

\begin{align*}
	%
	J(z)J(w)&\sim\OPE{2}{2(5+3k)}{},\\
	J(z)F^\pm(w)&\sim\pm\OPE{}{2}{}F^\pm(w),\\
	J(z)G^\pm(w)&\sim\pm\OPE{}{3}{}G^\pm(w),\\
	J(z)W^\pm(w)&\sim\pm\OPE{}{1}{}W^\pm(w),\\
	F^\pm(z)F^\pm(w)&\sim\,G^\pm(z)G^\pm(w)\sim F^\pm(z)G^\pm(w)\sim0,\\
	F^+(z)F^-(w)&\sim\OPE{2}{(5+3k)}{}+\OPE{}{1}{}J(w),\\
	G^+(z)G^-(w)&\sim\OPE{3}{2(4+3k)(5+3k)}{9}+\OPE{2}{(4+3k)}{3}J(w)\\
	+\OPE{}{1}{}\left(\right.&\left.-(4+k)L(w)+\frac{2}{3}\NO{F^+(w)F^-(w)}+\frac{1}{3}\NO{J(w)^2}+\frac{(2+3k)}{6}\partial J(w)\right),\\
	F^\pm(z)G^\mp(w)&\sim\OPE{}{1}{}W^\mp(w),\\
	W^\pm(z)W^\pm(w)&\sim\pm\OPE{2}{4(4+3k)}{3}F^\pm(w)\pm\OPE{}{2(4+3k)}{3}\partial F^\pm(w),\\
	W^+(z)W^-(w)&\sim-\OPE{3}{2(4+3k)(5+3k)}{3}-\OPE{2}{(4+3k)}{3}J(w)\\
	+\OPE{}{1}{}\left(\right.&\left.3(4+k)L(w)-\frac{10}{3}\NO{F^+(w)F^-(w)}-\frac{1}{3}\NO{J(w)^2}+\frac{(2-k)}{2}\partial J(w)\right),\\
	F^\pm(z)W^\pm(w)&\sim\OPE{}{3}{}G^\pm(w),\\
	F^\pm(z)W^\mp(w)&\sim\OPE{}{2}{}W^\pm(w),\\
	G^\pm(z)W^\pm(w)&\sim\pm\OPE{}{2}{3}\NO{F^\pm(w)^2},\\
	G^\pm(z)W^\mp(w)&\sim\mp\OPE{2}{2(4+3k)}{3}F^\pm(w)+\OPE{}{1}{}\left(-\frac{2}{3}\NO{J(w)F^\pm(w)}\mp\frac{(2+3k)}{3}\partial F^\pm(w)\right),\\
\end{align*}
where \[\charge=-\frac{2k(5+3k)}{4+k}.\]



The $\W$-algebra $\W^k(G_2,f_{\tilde{A_1}})$ is strongly generated by the fields $J(z)$, $F^\pm(z)$, $G^\pm(z)$, and $L(z)$ of conformal weights $1$, $1$, $\frac{5}{3}$ and $2$ respectively. They satisfy the OPEs:
\begin{align*}
	%
	J(z)J(w)&\sim\OPE{2}{(3+2k)}{},\\
	J(z)F^\pm(w)&\sim\pm\OPE{}{2}{}F^\pm(w),\\
	J(z)G^\pm(w)&\sim\pm\OPE{}{1}{}G^\pm(w),\\
	F^\pm(z)F^\pm(w)&\sim\,G^\pm(z)F^\pm(w)\sim0,\\
	F^+(z)F^-(w)&\sim\OPE{2}{(3+2k)}{2}+\OPE{}{1}{}J(w),\\
	F^\pm(z)G^\mp(w)&\sim\OPE{}{1}{}G^\pm(w),\\
	G^\pm(z)G^\pm(w)&\sim\mp\OPE{4}{2(2+k)(10+3k)(17+6k)}{}F^\pm(w)\mp\OPE{3}{(2+k)(10+3k)(17+6k)}{}\partial F^\pm(w)\\
	+\OPE{2}{1}{}&\left(\pm2(4+k)(16+5k)\NO{L(w)F^\pm(w)}\mp16(3+k)\NO{F^+(w)F^\pm(w)F^-(w)}\right.\\
	\mp&4(3+k)\NO{F^\pm(w)J(w)^2}-(44\mp48+(24\mp16)k+3k^2)\NO{F^\pm(w)\partial J(w)}\\
	+&\left.(2+k)(10+3k)\NO{J(w)\partial F^\pm(w)}\mp2(3+k)(38\mp4+20k+3k^2)\partial^2F^\pm(w)\right)\\
	+\OPE{}{1}{}&\left(\pm(4+k)(16+5k)\NO{L(w)\partial F^\pm(w)}\mp16(3+k)\NO{F^+(w)F^-(w)\partial F^\pm(w)}\right.\\
	\pm&(4+k)(16+5k)\NO{\partial L(w)F^\pm(w)}\mp8(3+k)\NO{F^\pm(w)^2\partial F^\mp(w)}\\
	\mp&4(3+k)\NO{F^\pm(w)J(w)\partial J(w)}\mp2(3+k)\NO{J(w)^2\partial F^\pm(w)}\\
	+&\frac{44+24k+3k^2}{2}(\NO{J(w)\partial^2F^\pm(w)}-\NO{F^\pm(w)\partial^2J(w)})\\
	\pm&\left.4(2\mp1)(3+k)\NO{\partial J(w)\partial F^\pm(w)}\mp\frac{940+736k+195k^2+18k^3}{12}\partial^3F^\pm(w)\right),\\
	G^+(z)G^-(w)&\sim\OPE{5}{(2+k)(3+2k)(10+3k)(17+6k)}{}+\OPE{4}{(2+k)(10+3k)(17+6k)}{}J(w)\\
	+\OPE{3}{1}{}&\left(-(4+k)(3+2k)(16+5k)L(w)+\frac{188+256k+119k^2+18k^3}{2}\partial J(w)\right.\\
	+&\left.(38+34k+7k^2)(4\NO{F^+(w)F^-(w)}+\NO{J(w)^2})\right)\\
	+\OPE{2}{1}{}&\left(-(4+k)(16+5k)\NO{L(w)J(w)}+8(3+k)\NO{F^+(w)F^-(w)J(w)}\right.\\
	+&(8+36k+11k^2)\NO{F^+(w)\partial F^-(w)} +(144+100k+17k^2)\NO{F^-(w)\partial F^+(w)}\\
	+&2(3+k)\NO{J(w)^3}+(26+30k+7k^2)\NO{J(w)\partial J(w)}\\
	-&\left.\frac{(4+k)(3+2k)(16+5k)}{2}\partial L(w)+(3+k)(42+20k+3k^2)\partial^2J(w)\right)\\
	+\OPE{}{1}{}&\left(\frac{3(4+k)^2}{2}\NO{L(w)^2}-\frac{3(2+k)(3+k)(4+k)}{2}\partial^2L(w)\right.\\
	-&\frac{(4+k)(16+5k)}{2}\NO{\partial L(w)J(w)}-\frac{(4+k)(8+5k)}{2}\NO{L(w)\partial J(w)}\\
	+&\frac{396+332k+90k^2+9k^3}{12}\partial^3J(w)-8(4+k)\NO{L(w)F^+(w)F^-(w)}\\
	-&2(4+k)\NO{L(w)J(w)^2}+\frac{116+72k+15k^2}{2}\NO{\partial F^+(w)\partial F^-(w)}\\
	+&\frac{(2+k)(26+15k)}{8}\NO{\partial J(w)^2}+\frac{96+52k+9k^2}{4}\NO{J(w)\partial^2 J(w)}\\
	+&2(27+19k+3k^2)\NO{F^-(w)\partial^2F^+(w)}+(26+14k+3k^2)\NO{F^+(w)\partial^2F^-(w)}\\
	+&(7+3k)\NO{J(w)^2\partial J(w)}+4(5+k)\NO{J(w)F^-(w)\partial F^+(w)}\\
	+&4(1+k)\NO{J(w)F^+(w)\partial F^-(w)}+4(k-1)\NO{F^+(w)F^-(w)\partial J(w)}\\
	+&\left.8\NO{F^+(w)^2F^-(w)^2}+4\NO{F^+(w)F^-(w)J(w)^2}+\frac{1}{2}\NO{J(w)^4}\right),
\end{align*}
where \[\charge=-\frac{(92+81k+18k^2)}{4+k}.\]



The $\W$-algebra $\W^k(G_2,f_{\subreg})$ is of type $(2^3,3)$ strongly generated by the fields $L(z)$, $G^\pm(z)$ and $F(z)$. They satisfy the OPEs:
\begin{align*}
	%
	G^+(z)F(w)&\sim\OPE{3}{2(2 + k) (16+5k)}{}G^-(w)+\OPE{2}{(2 + k) (16+5k)}{2}\partial G^-(w)\\
	&+\OPE{}{1}{}\left(2\NO{G^+(w)G^-(w)}-2(4+k)\NO{L(w)G^-(w)}\right.\\
	&\qquad\left.+2\partial F(w)+\frac{(2+k)^2}{2}\partial^2 G^-(w)\right),\\
	G^-(z)F(w)&\sim\OPE{3}{2(2+k)(16+5k)}{}G^+(w)+\OPE{2}{(2+k)(16+5k)}{2}\partial G^+(w)\\
	&+\OPE{}{1}{}\left(-\NO{G^+(w)^2}-2(4+k)\NO{L(w)G^+(w)}\right.\\
	&\qquad\left.-\NO{G^-(w)^2}+\frac{(2+k)^2}{2}\partial^2G^+(w)\right),\\
	F(z)F(w)&\sim-\OPE{6}{(2 + k) (10 + 3 k) (16 + 5 k)(4+k)c_k}{2}\\
	&-\OPE{4}{3(2+k)(4+k)(10+3k)(16+5k)}{}L(w)\\
	&-\OPE{3}{3(2+k)(4+k)(10+3k)(16+5k)}{2}\partial L(w)\\
	&+\OPE{2}{1}{}\left(-(8+3k)\NO{G^+(w)^2}+2(4 + k)^2 (10+3k)\NO{L(w)^2}\right.\\
	&\left.+(8+3k)\NO{G^-(w)^2}-\frac{3(2+k)(4+k)(8+3k)(10+3k)}{4}\partial^2 L(w)\right)\\
	&+\OPE{}{1}{}\left(-(8+3k)\NO{G^+(w)\partial G^+(w)}+2(4+k)^2(10+3k)\NO{L(w)\partial L(w)}\right.\\
	&\left.+(8+3k)\NO{G^-(w)\partial G^-(w)}-\frac{(2 + k) (4 + k) (4 + 3 k) (10 + 3 k)}{6}\partial^3 L(w)\right),\\
	G^\pm(z)G^\pm(w)&\sim\pm\OPE{4}{(10+3k)(4+k)c_k}{2}+\OPE{2}{1}{}\left(\pm2(4 + k) (10 + 3 k)L(w)-4(3+k)G^+(w)\right)\\
	&+\OPE{}{1}{}\left(\pm(4 + k) (10 + 3 k)\partial L(w)-2(3+k)\partial G^+(w)\right),\\
	G^+(z)G^-(w)&\sim\OPE{2}{4(3+k)}{}G^-(w)+\OPE{}{1}{}\left(-2F(w)+2(3+k)\partial G^-(w)\right),
\end{align*}
where \[\charge=-\frac{4(k+2)(17+6k)}{4+k}.\]


To our knowledge, OPEs of the principal $\W$-algebra $\W^k(G_2)$ were computed first in \cite{MPS22}. For completeness of the paper, we recall them in this section.
%
%
The $\W$-algebra $\W^k(G_2)$ is of type $(2,6)$, strongly generated by the fields $L(z)$ and $W(z)$, and satisfies the OPE:
\begin{align*}
%
W(z)W(w)&\sim
\OPE{12}{-w_{0,k}c_k}{81} 
+\OPE{10}{2w_{0,k}}{27} L(w)
+\OPE{9}{w_{0,k}}{27} \partial L(w)\\
+&\OPE{8}{w_{1,k}}{108}\left(-62 (k+4)\NO{L(w)^2}+3 \left(84 k^2+579 k+1000\right)\partial^2L(w)\right)\\
+&\OPE{7}{w_{1,k}}{108}\left(-62 (k+4)\NO{L(w)\partial L(w)}+\frac{2}{3} \left(84 k^2+579 k+1000\right)\partial^3L(w)\right)\\
+&\OPE{6}{1}{}(2w_{3,k}W(w)+w_{2,k}\Lambda^5(L))+\OPE{5}{1}{}(w_{3,k}\partial W(w)+w_{2,k}\Lambda^4(L))\\
+&\OPE{4}{1}{}(w_{5,k}\left(\frac{10}{9} \left(588 k^2+3929 k+6504\right)\partial^2W(w)-\frac{1240}{3} (k+4)\NO{L(w)W(w)}\right)\\
&\quad+w_{4,k}\Lambda^3(L))\\
+&\OPE{3}{1}{}(w_{5,k}\Omega^2(W,L)+w_{4,k}\Lambda^2(L))+\OPE{2}{1}{}(w_{7,k}\Omega^1(W,L)+w_{6,k}\Lambda^1(L))\\
+&\OPE{}{1}{}(w_{7,k}\Omega^0(W,L)+w_{6,k}\Lambda^0(L)),
\end{align*}
where 
\[\charge=-\frac{2(12k+41)(7k+24)}{4+k}.\]
and 
\begin{align*}
	w_{0,k}=&2(k+4)^2 (2 k+5) (2 k+7) (3 k+10) (7 k+22) (7 k+23) (8 k+27) (9 k+34)\\ &(11 k+40) (12 k+37) (15 k+52) (15 k+53) (18 k+65)(336 k^2+2301 k+3940)\\&(588 k^2+3991 k+6752))\\
	w_{1,k}=& (k+4)^2 (2 k+5) (2 k+7) (3 k+10) (7 k+22) (8 k+27) (9 k+34) (11 k+40)\\ & (12 k+37) (15 k+52) (18 k+65) \left(336 k^2+2301 k+3940\right) \left(588 k^2+3991k+6752\right)\\
	w_{2,k}=&(k+4)^2 (2 k+5) (2 k+7) (3 k+10) (7 k+22) (8 k+27) (9 k+34) (11 k+40)\\ & (12 k+37) (15 k+52) (18 k+65)\\
	w_{3,k}=&\frac{140}{9}(k+4) (2 k+7) (3 k+10) (7 k+20) (12 k+35) (13 k+48) (24 k+89)\\& \left(3 k^2+24 k+47\right) \left(336 k^2+2301 k+3940\right) \left(588 k^2+3991 k+6752\right)\\
	w_{4,k}=&(k+4)^2 (2 k+5) (2 k+7) (3 k+10) (9 k+34) (11 k+40) (12 k+37)\\
	w_{5,k}=&(k+4) (2 k+7) (3 k+10) (7 k+20) (24 k+89) \left(3 k^2+24 k+47\right)\\& \left(336 k^2+2301 k+3940\right) \left(588 k^2+3991 k+6752\right)\\
	w_{6,k}=&(k+4)^2 (2 k+5) (2 k+7) (3 k+10) (9 k+34)\\
	w_{8,k}=&(k+4) \left(3 k^2+24 k+47\right) \left(336 k^2+2301 k+3940\right) \left(588 k^2+3991 k+6752\right),
\end{align*}
The fields $\Lambda^n(L)$ only depend on $L$ while every summand in the fields $\Omega^n(W,L)$ depends on $W$. We refer to \cite[Appendix]{MPS22} for their explicit definition.

\section{Isomorphisms of $\W$-algebras}\label{sec:collapsing_levels}

In this section, we use the explicit description of $\W$-algebras associated with Lie algebras of rank two to obtain remarkable isomorphisms of vertex algebras (Table~\ref{table:collapse}).
%
%

\subsection{Trivial simple $\W$-algebras}

Consider the $\W$-algebra $\WA$ associated with a nilpotent element $f\in\g$. Denote $\Imax$ its maximal ideal -- which depends on the level $k$. We have $\SWA=\WA/\Imax$.
At non-critical level, it follows from the conformal structure of $\WA$ that the vertex algebra $\SWA$ is trivial, i.e. isomorphic to $\C\vacuum$, if and only if the conformal vector generates $\Imax$.
This happens only if the central charge $\charge$ is zero, but this necessary condition is not sufficient.
For instance, the conformal vector $L$ of $\W_k(\sp_4,\fmin)$ does not belong to the maximal ideal when $k=-1$ whereas $c_{-1}=0$.
Indeed, according to the OPEs described in Sect.~\ref{subsec:OPEsB2}, if $L$ were in $\mc{I}_{-1}$ then $J$ would be in $\mc{I}_{-1}$ as well as the vacuum.


Adding a supplementary condition, we obtain a complete criterion. We enjoy this opportunity to mention that the proofs of Propositions~2.2.2 and 2.2.7 in \cite{Fasthesis} are incomplete. We correct them in this paper. It does not impact the results presented in the thesis.

\begin{Proposition}\label{prop:criterium_trivial_SWA}
	Let $J^{(1)},\ldots, J^{(n)}$ be a set of strong generators of $\WA$ with conformal weights $w_1\ldots w_n\in\R_{>0}$ respectively. 
	The simple quotient $\SWA$ is trivial if and only if for any $1\leq i,j\leq n$ 
	the pole of degree $w_i+w_j$ of the OPE $J^{(j)}(z)J^{(i)}(w)$ vanishes, that is $(J^{(j)})_{w_i}J^{(i)}=0$.
\end{Proposition}

\begin{Proof}
	Since the $\W$-algebra $\WA$ is positively graded by the conformal vector, its simple quotient is trivial if and only if all the strong generators are in the maximal ideal $\Imax$.
	For all $1\leq i,j\leq n$, the vector $(J^{(j)})_{w_i}J^{(i)}$ has conformal weight zero. Since $\WA$ is canonical, i.e. $\WA_0=\C\vacuum$, it vanishes.
	
	Conversely, let $1\leq i,j\leq n$. For $m>0$, $(J^{(j)})_mJ^{(i)}$ has conformal weight $w_i-m$ and it is a linear combination of normally ordered products and derivatives of strong generators $J^{(\ell)}$ of conformal weight $w_\ell$ lower than $w_i-m<w_i$. In particular, $(J^{(j)})_{w_i}J^{(i)}$ is a multiple of the vacuum: it exists $c_{i,j}\in\C$ such that
	$$(J^{(j)})_{w_i}J^{(i)}=c_{i,j}\vacuum.$$
	It  follows that $J^{(i)}$ is in $\Imax$ if all strong generators $J^{(\ell)}$ of conformal weight lower than $w_i$ are in $\Imax$ and if $c_{i,j}=0$ for all $1\leq j\leq n$.
	Note that if $J^{(i)}$ has the lowest conformal weight, $(J^{(j)})_mJ^{(i)}=0$ for all $m\neq w_i$ and so $J^{(i)}$ is in $\Imax$ if and only if $(J^{(j)})_{w_i}J^{(i)}=0$.
	
	Therefore, if $(J^{(j)})_{w_i}J^{(i)}=0$ for all $1\leq i,j\leq n$, then all the strong generators are in $\Imax$ and thus $\SWA\simeq\C\vacuum$.
	
	

\end{Proof}



We obtain the following classification.

\begin{Corollary}\label{cor:trivial_collapse}
	\begin{enumerate}
		\item Consider the Lie algebra $\g=\sl_3$, then $\SWA\simeq\C$ if and only if
		\[(f,k)\in\{(\fmin,-3/2), (f_{\reg},-5/3), (f_{\reg},-9/4)\}.\]
		\item Consider the Lie algebra $\g=\sp_4$, then $\SWA\simeq\C$ if and only if
		$$(f,k)\in\{{(\fmin,-1/2)},(f_{\reg},-13/6),(f_{\reg},-12/5)\}.$$
		\item Consider the Lie algebra $\g=G_2$, then $\SWA\simeq\C$ if and only if
		$$(f,k)\in\{{(\fmin,-5/3)},(f_{\subreg},-17/6),(f_{\subreg},-2),(f_\reg,-41/12),(f_\reg,-24/7)\}.$$
	\end{enumerate}
\end{Corollary}


Most of the pairs listed in the Corollary~\ref{cor:trivial_collapse} have already been established in \cite{AKM+18,AvEM22}. However, we get new isomorphisms such as the one involving $\W_{-2}(G_2,f_\subreg)$. It provides a new example of rational $\W$-algebra at non admissible level.

\begin{Corollary}\label{cor:G2rational}
	The simple $\W$-algebra $\W_{-2}(G_2,f_\subreg)$ is $C_2$-cofinite and rational.
\end{Corollary}


The principal $\W$-algebras $\W_k(\sl_3)$, $\W_k(\sp_4)$ and $\W_k(G_2)$, and the subregular $\W$-algebra $\W_k(G_2,f_{\subreg})$ are one-dimensional if and only if $c_k=0$. More generally, we conjecture the following:

\begin{Conjecture}\label{conj:collaps_distinguished}
	If $f$ is distinguished, i.e. $\g^f_0=0$, then $\SWA\simeq\C\vacuum$ if and only if $c_k=0$.
\end{Conjecture}

We check this conjecture in many cases. In particular, we verify in \cite{AvEM22} that, when $f$ is distinguished and $k$ is an admissible level which vanishes the central charge, $\SWA\simeq\C\vacuum$.

In another direction, the Corollary~\ref{cor:G2rational} gives more evidence for the following conjecture (see \cite[Chap.~6.3]{Fasthesis} for additional details). We will explore in depth this topic in our futur works.

\begin{Conjecture}\label{conj:AV_G2}
	The associated variety of the simple vertex algebra $L_{-2}(G_2)$ is the closure of the subregular nilpotent orbit of $G_2$:
	$$X_{L_{-2}(G_2)}=\overline{\O}_{\subreg}.$$
\end{Conjecture}

%

\subsection{Simple quotients isomorphic to Virasoro vertex algebras}

From now on, we consider non-trivial simple $\W$-algebras. In particular, the conformal vector does not belong to the maximal ideal $\Imax$ of $\WA$. If the other strong generators are in $\Imax$, then $\SWA$ is isomorphic to the Virasoro vertex algebra $\Vir_{c_k}$ with central charge $c_k$.

\begin{Proposition}
	Let $J^{(1)},\ldots, J^{(n)}:=L$ be a set of strong generators of $\WA$ with conformal weights $w_1\ldots w_n\in\R$ respectively. 
	The simple quotient $\SWA$ is isomorphic to the simple Virasoro vertex algebra $\Vir_{c_k}$ if and only if for any $1\leq i\leq n-1$ and $1\leq j\leq n$, $(J^{(j)})_{w_i}J^{(i)}=0$ and $(J^{(j)})_{w_i-2}J^{(i)}$ does not depend on $L$.	
\end{Proposition}


\begin{Proof}
	The proof is similar to the one of Proposition~\ref{prop:criterium_trivial_SWA}.
	The simple quotient $\SWA$ is isomorphic to $\Vir_{c_k}$ if and only if it is strongly generated only by the conformal vector $L$.
	Hence, $\SWA\simeq\Vir_{c_k}$ if and only if $J^{(1)},\ldots, J^{(n-1)}\in\Imax$ and $L\notin\Imax$.
	Since $J^{(1)},\ldots, J^{{(n-1)}}\in\Imax$ , then $(J^{(j)})_{w_i}J^{(i)}=0$ for all $1\leq i\leq n-1$ and $1\leq j\leq n$.  Moreover,
	$(J^{(j)})_{w_i-2}J^{(i)}$ is a linear combination of normally ordered products and derivatives of strong generators $J^{(\ell)}$ of conformal weight lower than $2$. There exist $c_{i,j}\in\C$ and a linear combination $R_{i,j}$ of normally ordered products and derivatives of the strong generators $J^{(\ell)}$, $1\leq\ell\leq n-1$, such that
	\begin{equation}\label{eq:decompo2}
		(J^{(j)})_{w_i-2}J^{(i)}=c_{i,j}L+R_{i,j}.
	\end{equation}
	Since $(J^{(j)})_{w_i-2}J^{(i)}$ and $R_{i,j}$ belongs to $\Imax$, we deduce that $c_{i,j}=0$. Thus, $(J^{(j)})_{w_i-2}J^{(i)}$ does not depend on $L$.	
	
	Conversely, let $1\leq i\leq n-1$ and $1\leq j\leq n$. For $m>0$, $(J^{(j)})_mJ^{(i)}$ is a sum of normally ordered products and derivatives of strong generators $J^{(\ell)}$ of conformal weight lower than $w_i-m<w_i$. 
	In particular, $(J^{(j)})_{w_i}J^{(i)}$ is a multiple of the vacuum. In addition, $L$ can appear only as a summand in the decomposition of $(J^{(j)})_{w_i-2}J^{(i)}$. We have a decomposition \eqref{eq:decompo2}.
	Since $L\notin\Imax$, $J^{(i)}$ is in the maximal ideal if $(J^{(j)})_{w_i}J^{(i)}=0$ and $(J^{(j)})_{w_i-2}J^{(i)}$ does not depend on $L$ for all $1\leq j\leq n$.
	Therefore, if this condition is satisfied for all $1\leq i\leq n-1$, then all the strong generators except $L$ are in $\Imax$ and $\SWA\simeq\Vir_{c_k}$.
%
\end{Proof}


We deduce the following classification from the OPEs presented in Sect.~\ref{sec:OPEs}.

\begin{Corollary}\label{cor:list_vir}
	Consider the simple Lie algebra $\g$ of rank $2$.
	\begin{enumerate}
		\item The simple $\W$-algebra $\W_k(\sl_3)$ is isomorphic to $\Vir_{c_k}$ if and only if $k=-4/3$ or $k=-12/5$.
		\item If $\g=\sp_4$, then $\SWA\simeq\Vir_{c_k}$ if and only if
		\begin{itemize}
			\item $f=f_{\subreg}$ and $k=-2$, or
			\item $f=f_{\reg}$ and $k\in\{-3/2,-8/3,-11/5,-11/6,-18/7,-19/8\}$.
		\end{itemize}
		\item If $\g=G_2$, then $\SWA\simeq\Vir_{c_k}$ if and only if
		\begin{itemize}
			\item $f=f_{\tilde{A_1}}$ and $k=-3/2$, or
			\item $f=f_{\subreg}$ and $k=-10/3$, or
			\item \cite[Proposition~5.3]{MPS22} $f=f_{\reg}$ and $$k\in\{-\tfrac{5}{2},-\tfrac{7}{2},-\tfrac{10}{3},-\tfrac{22}{7},-\tfrac{23}{7},-\tfrac{27}{8},-\tfrac{34}{9},-\tfrac{40}{11},-\tfrac{37}{12},-\tfrac{52}{15},-\tfrac{53}{15},-\tfrac{65}{18}\}.$$
		\end{itemize}
	\end{enumerate}
\end{Corollary}


The OPEs of $\W^k(G_2,f_\subreg)$ also admit simplifications at level $k=-16/5$ (see Sect.~\ref{subsec:OPEsG2}). Indeed, the strong generator $F$ belongs to the maximal ideal $\mc{I}_{-16/5}$. This induces relations between the three other strong generators and their derivatives in the simple $\W$-algebra.
Using the OPEs, the authors of \cite{MPS22} proved that the simple $\W$-algebra corresponds with the tensor product of three copies of the Virasoro vertex algebra of central charge $-22/5$ see 
\begin{equation*}
\W_{-16/5}(G_2,f_\subreg)\simeq(\Vir_{-22/5})^{\*3}.
\end{equation*}
Since $\Vir_{-22/5}$ is $C_2$-cofinite and rational \cite{Ara12}, the tensor product $\W_{-16/5}(G_2,f_\subreg)$ is also $C_2$-cofinite and rational \cite{ABD04, DMZ94}.


\subsection{Other isomorphisms}

For several additional particular simple $\W$-algebras, the OPEs give a complete or partial description of the maximal ideal.
For instance, we recover some isomorphisms from \cite{AKM+17} and \cite[Theorem~10.10]{AvEM22}:
\begin{align*}
	&\W_{-2}(\sp_4,\fmin)\simeq M(1),&&\W_{-2}(\sp_4,\fmin)\simeq L_{-3/2}(\sl_2),\\
	&\W_{-4/3}(G_2,\fmin)\simeq L_1(\sl_2),&&\W_{-17/6}(G_2,f_{\tilde{A}_1})\simeq L_{-4/3}(\sl_2),
\end{align*}
where $M(1)$ denotes the Heisenberg vertex algebra of central charge $1$ and $L_k(\g)$ is the simple affine vertex algebra associated with $\g$ at level $k$.
We are also able to determine other isomorphisms of vertex algebras.

\begin{Proposition}\label{prop:new_collapsing}
	We have the following isomorphisms:
	\begin{gather*}
		\W_{-10/3}(G_2,f_{\tilde{A_1}})\simeq L_{-11/6}(\sl_2),\qquad\W_{-2}(G_2,f_{\tilde{A_1}})\simeq L_{-\frac{1}{2}}(\sl_2),\\
		\W_{-1}(\sp_4,f_{\subreg})\simeq M(1).
	\end{gather*}
\end{Proposition}

\begin{Proof}
	We detail the proof of the isomorphism $\W_{-1}(\sp_4,f_{\subreg})\simeq M(1)$. Other cases are similar.
	Using the OPEs, we check that when $k=-1$, the vector $G^+$ and $G^-$ generate a proper ideal. Hence, they belongs to the maximal ideal $\W^{-1}(\sp_4,f_{\subreg})$.
	It follows that $G^+_{(n)}G^-=0$ for all $n\geq0$. In particular, $G^+_{(1)}G^-=0$ gives  a relation between the conformal field $L(z)$ and the field $\NO{J(z)^2}$
	$$L(z)=\frac{1}{2}\NO{J(z)^2}.$$
	Thus, $L$ coincides with the conformal vector of the Heisenberg vertex algebra generated by $J$.
	Since $J$ does not belong to the maximal ideal of $\W^{-1}(\sp_4,f_{\subreg})$, we get the isomorphism.
\end{Proof}

We summarize in Table~\ref{table:collapse} the isomorphisms of $\W$-algebras associated with simple Lie algebras of rank two we mentioned in Sect.~\ref{sec:collapsing_levels}.
\begin{table}[h!]
	\centering
	\renewcommand{\arraystretch}{1.3}
	\begin{tabular}{|c||M{3.2cm}|c||c|M{2.4cm}|}
		\hline
		$\W=\SWA$&affine vertex &$k$ such that  &$\mathcal{V}$&$k$ such that \\
		&algebra $L=L_{k'}(\g')$&$\W\simeq L$&&$\W\simeq \mathcal{V}$\\
		\hline\hline
		$\W_k(\sl_3,\fmin)$&$M(1)$& $-1$ &$\C$&$-\frac{3}{2}$ \\
		\hline
		$\W_k(\sl_3)$&$\C$&$-\frac{5}{3}$, $-\frac{9}{4}$&$\Vir_{c_k}$&$-\frac{4}{3}$, $-\frac{12}{5}$\\
		\hline
		$\W_k(\sp_4,\fmin)$&$L_{-3/2}(\sl_2)$&$-2$&$\C$&$-\frac{1}{2}$\\
		\hline
		$\W_k(\sp_4,f_\subreg)$&$M(1)$&$-1$&$\Vir_{c_k}$&$-2$\\
		\hline
		$\W_k(\sp_4)$&$\C$&$-\frac{13}{6}$, $-\frac{12}{5}$&$\Vir_{c_k}$&$-\frac{3}{2}$, $-\frac{8}{3}$, $-\frac{11}{5}$, $-\frac{11}{6}$, $-\frac{18}{7}$, $-\frac{19}{8}$\\
		\hline
		$\W_k(G_2,\fmin)$&$L_{1}(\sl_2)$&$-\frac{4}{3}$&$\C$&$-\frac{5}{3}$\\
		\hline
		$\W_k(G_2,f_{\tilde{A_1}})$&$L_{p(k)}(\sl_2)$ with $p(k)=-\frac{92+81 k+18 k^2}{52 + 42 k + 9 k^2}$&$-\frac{17}{6}$, $-\frac{10}{3}$, $-2$ &$\Vir_{c_k}$& $-\frac{3}{2}$  \\
		\hline
		$\W_k(G_2,f_\subreg)$&$\C$&$-\frac{17}{6}$, $-2$&$\Vir_{c_k}$&$-\frac{10}{3}$\\
		\hline
		$\W_k(G_2)$&$\C$&$-\frac{41}{12}$, $-\frac{24}{7}$ &$\Vir_{c_k}$&$-\frac{5}{2}$, $-\frac{7}{2}$, $-\frac{10}{3}$, $-\frac{22}{7}$,
		$-\frac{23}{7}$, $-\frac{27}{8}$, $-\frac{34}{9}$,  $-\frac{40}{11}$,
		$-\frac{37}{12}$, $-\frac{52}{15}$, $-\frac{53}{15}$, $-\frac{65}{18}$\\
		\hline
	\end{tabular}
	\caption{Isomorphisms for $\W$-algebras of rank two.}\label{table:collapse}
\end{table}

\begin{Remark}
	In the Table~\ref{table:collapse}, the levels appearing in the third column are called \emph{collapsing} \cite{AKM+17,AKM+18,AvEM22}.
	The simple affine vertex algebra $L_{k'}(\g')$ in the second column corresponds to $L_{k^\natural}(\g^\natural)$ where $\g^\natural$ is the centralizer of the $\sl_{2}$-triple $\{f,h,e\}$ in $\g$ and $k^\natural$ is defined as in \cite{KW04}. 
\end{Remark}

\printbibliography

\end{document}